\documentclass[showpacs,amssymb,floatfix,pre,aps,notitlepage]{revtex4}

\usepackage{amsfonts,bm}
\usepackage{graphics,graphicx}
\usepackage{amsmath}
\usepackage{color}

\newcommand{\Zbare}{Z _{\text{bare}}}
\newcommand{\eff}{\text{eff}}
\newcommand{\bare}{\text{bare}}
\newcommand{\sat}{\text{sat}}
\newcommand{\salt}{\text{salt}}
\newcommand{\res}{\text{res}}
\newcommand{\back}{\text{back}}
\newcommand{\wt}{\widetilde}

\begin{document}

\title{The renormalized jellium model for colloidal mixtures}
\author{Mar\'ia Isabel Garc\'ia de Soria}
\affiliation{F\'{\i}sica Te\'{o}rica, Universidad de Sevilla,
Apartado de Correos 1065, E-41080, Sevilla, Spain}
\author{Carlos E. \'Alvarez}
\affiliation{Facultad de Ciencias Naturales y Matem\'aticas, Universidad del Rosario, Calle 12C No. 6-25, Bogot\'a, Colombia}
\author{Emmanuel Trizac}
\affiliation{LPTMS, CNRS, Univ. Paris-Sud, Universit\'e Paris-Saclay, 91405 Orsay, France}

\begin{abstract}
In an attempt to quantify the role of polydispersity in colloidal suspensions,
we present an efficient implementation of the renormalized jellium model for a mixture of
spherical charged colloids. The different species may have different size, charge
and density. Advantage is taken from the fact that 
the electric potential pertaining to a given species obeys a Poisson's equation 
that is species independent; only boundary conditions do change from a species
to the next. All species are coupled through the renormalized background (jellium)
density, that is determined self-consistently. The corresponding predictions are
compared to the results of Monte Carlo simulations of binary mixtures, where 
Coulombic interactions are accounted for exactly, at the primitive model level
(structureless solvent with fixed dielectric permittivity).
An excellent agreement is found.
\end{abstract}
\maketitle

\section{Introduction}
\label{sec:intro}

Predicting structural and thermodynamic properties of charged colloidal
suspensions is a difficult task \cite{Belloni,HL00,Levin,Messina09}.
At the simplest level of description, the solvent is treated as a continuous
medium of fixed dielectric permittivity and one discards correlation effects
that prevail, as a rule of thumb, for multivalent micro-ions and sufficiently 
charged colloids \cite{Levin}. Viewing  the microionic fluid as an inhomogeneous ideal gas 
leads to the Poison-Boltzmann theory. However, as such, it does not easily lend itself
to numerical investigations \cite{Fushiki,Dobnikar}, not to mention analytical progress.
In practice, this mean-field theory often needs a further mean-field-like reduction,
to predict quantities that can be compared to experiments or simulations, such as osmotic
pressures. One successful and popular such simplification is the so-called cell model,
where an $N$-body colloidal situation is mapped onto a one body problem, placed at the center
of a Wigner-Seitz cell \cite{Marcus}. This cell is often taken spherical for simplicity,
with a volume equal to the mean volume per colloid. As an alternative to the cell picture, 
a renormalized jellium model was proposed in Ref. \cite{tl04}, elaborating on an idea put forward 
by Beresford-Smith {\it et al.} \cite{Beresford}, who nevertheless did not implement the renormalization 
procedure, which turns out crucial \cite{tl04,ptl07,castaneda1,castaneda2}. 

For monodisperse colloidal suspensions, both cell and jellium models yield very close and accurate results 
for quantities that can be compared against numerical simulations or experiments \cite{levin_tb03,tl04,NJP2006,Denton1,Denton2}. 
Yet, when it comes to colloidal mixtures, the cell model is not free of ambiguities 
\cite{ttvr08}, whereas the jellium model admits a natural extension \cite{castaneda1,castaneda2}. In light of the intrinsic interest 
in polydisperse suspensions \cite{Krause91,hd05,hcvrvbd06,ywy12}, our goal here is three-fold. 
First, we present in section \ref{sec:method} the main ingredients of the jellium model,
together with a new procedure that allows to solve the problem self-consistently for
mixtures, in a more efficient way than hitherto proposed. Compared to the  
method used in Refs \cite{tl04,ptl07} for monodisperse colloids, an elegant reformulation 
was reported in \cite{castaneda1,castaneda2}, that significantly speeds up the resolution. We shall argue that 
this reformulation looses its suitability when dealing with mixtures. 
Second, we discuss in section \ref{sec:results}
some of the main features of effective charges as emerging within the jellium
approach. Yet, such quantities, interesting in their own right, 
can be coined as 'secondary', in the sense that they are often not directly measured
in an experiment or in a simulation. We therefore implement 
Monte Carlo simulations of a binary charged mixture, which provide an important benchmark against 
which the polydisperse cell and the jellium schemes can be confronted. Our simulations,
at the level of the primitive model, do not rely an any mean-field hypothesis,
and treat exactly the Coulombic nature of  the interactions between all species
(colloids and micro-ions). Conclusions are finally drawn in section \ref{sec:concl}.

For the following discussion, it seems appropriate to revisit briefly an aspect of
the common phenomenology of cell and jellium effective charges. For highly charged
colloids [yet in the mean-field regime, where a Poisson-Boltzmann description may hold],
the strong interactions between the colloids and the micro-ions induce an accumulation 
of the latter in the vicinity of the colloids. This in turn induces a renormalization 
of the colloidal effective charge \cite{Belloni,Levin,alex,tba02,tbavg03}. 
If the colloidal bare charge $Z_\bare$ is large, the effective charge become independent 
of $Z_\bare$; this is the saturation phenomenon \cite{TT03}, a signature of mean-field, where the effective charge becomes
$Z_\sat$, which only depends on the density and salt content.
For a reason to become clear below, in the no salt case, $Z_\sat$ as a function of density (or volume fraction $\eta$) exhibits a non monotonous 
behavior, very close to that of the function $f(X,\infty$) versus $X$ in Fig. \ref{fig:f}. For 
small $\eta$ (equivalently, small $X$ in Fig. \ref{fig:f}), the effective charge decreases 
with increasing $\eta$. This is an entropy effect, whereby a lowering of $\eta$ induces a dilution 
of micro-ions, which leave the vicinity of the colloids to gain translational entropy \cite{Palberg_Lowen14}.
In other words, increasing $\eta$, less volume is available for the microions, electrostatic 'binding' 
is stronger, and $Z_\sat$ consequently decreases. However, further increasing $\eta$,
$Z_\sat$ starts to increases: this can be viewed as an indirect effect of screening.
The micro-ions efficiently screen their own interactions with the colloids, 
so that electrostatic binding is weakened. This dichotomy between the entropy dominated and the energy
dominated regimes will be met again below, where it induces a non-trivial dependence on mixture
composition.

\section{The renormalized jellium: principles and resolution}
\label{sec:method}

\subsection{A (mean-field)$^2$ approach}
\label{ssec:nosalt}

We consider an arbitrary mixture of positively charged spherical colloids, where each species is indexed by 
an integer $i$. The radius of species $i$ having number density $\rho_i$ is  
$a_i$, while $e Z^i_{\text bare}$ stands for the bare charge, $e>0$ being the elementary charge. 
The total density is $\rho=\sum_i \rho_i$, and to characterize the composition of the mixture, it is convenient to introduce the 
molar fraction $x_i = \rho_i/\rho$, such that $\sum_i x_i=1$.
The starting point of the jellium 
model is the same as the celebrated Poisson-Boltzmann theory \cite{Belloni,Levin}, with an additional
assumption, that allows to restrict the problem to a single colloid
formulation (the cell model approach also aims at a similar restriction,
but proceeds very differently \cite{ttvr08}). 
The key point in the jellium approach is that the charge of other colloids
around a given tagged macroion is smeared out to form a homogeneous background
of charge density $\rho \,e Z_{\text back}$, in which the small ions are then
immersed. 
A self-consistency requirement connects this background charge
with the effective charge of the various species, see Refs \cite{tl04,ptl07,castaneda1,castaneda2}
for more details.

We denote the Bjerrum length by $\ell_B$, and we restrict for the sake
of the argument to salt free systems (see subsection \ref{ssec:salt} for the general case). 
The dimensionless electrostatic potential around a given colloid of type $i$,
centered at position $\mathbf{r}=0$ 
then obeys \cite{tl04,ptl07,castaneda1,castaneda2}
\begin{equation}
\nabla^2 \phi_i \,=\, 4 \pi \ell_B \,\rho Z_{\text back} \, \left(e^{\phi_i}-1 \right)
\label{eq:Poisson_nosalt}
\end{equation}
with boundary conditions
\begin{equation}
\phi_i \to 0 \hbox{ ~ for ~ } r \equiv |\mathbf{r}| \to \infty \quad \hbox{ ~ and ~ } \quad 
\frac{d \phi_i}{dr} = - Z^i_{\text bare} \ell_B \, \frac{1}{a_i^2}\hbox{ ~ at ~ } r=a_i.
\label{eq:bc}
\end{equation}

The first contribution on the r.h.s. stems from the counter-ions and takes the usual 
Poisson-Boltzmann form, while the second
is that of the smeared out background. 
Self-consistency demands that \cite{tl04}
$\rho Z_{\back} = \sum_i \rho_i Z^i_{\text{eff}},$
where $Z^i_{\text{eff}}$ is the effective charge of species $i$, defined 
from the far-field (large $r$) behavior of $\phi_i$ \cite{tba02}.
Since all species obey the same differential equation, but with different
boundary conditions, it follows that their effective charge is given by a 
unique two-parameter function $f$
\begin{equation}
\frac{Z^i_{\eff} \,\ell_B}{a_i} \,=\, f\left(X_i, \frac{Z^i_{\bare}\,\ell_B}{a_i}\right),
\label{eq:blabla}
\end{equation}
where 
\begin{equation}
X_i = 4 \pi \ell_B \rho Z_{\back} a_i^2 /3
\end{equation}
 is a dimensionless parameter, that 
scales like $a_i^2$ from one species to the next. The reason for including the factor $3$
in the definition of $X_i$ will become clear below.
Once the function $f(X,Y)$ is known,
the self-consistency condition determines $Z_\back$:
\begin{equation}
Z_\back  \,=\, \sum_i x_i \, Z^i_{\text{eff}} \,=\,
\frac{1}{\ell_B} \sum_i x_i\, a_i\, f\left(4 \pi \ell_B \rho Z_{\back} a_i^2/3,\frac{Z^i_{\bare}\,\ell_B}{a_i}\right).
\label{eq:selfcons}
\end{equation}
At this stage, it can be appreciated that the renormalized jellium model
is a mean-field simplification of an otherwise mean-field  (Poisson-Boltzmann) starting point.
The $N$-body Poisson-Boltzmann problem is a notoriously difficult problem
to solve from a computational viewpoint (not speaking of the lack of
analytical results) \cite{Dobnikar}.
With the renormalized jellium, a complex mixture problem is mapped onto a series of single colloid
equations (\ref{eq:Poisson_nosalt}), in a common background with density $\rho Z_\back$ to which all species contribute
(see Eq. (\ref{eq:selfcons})), acting thereby as a coupling term.

\subsection{Self consistent resolution}

In the subsequent analysis, we will single out species 1, and use its radius $a_1$ as our reference
length scale. Since colloidal charges appear in conjunction with the ratio $\ell_B$ over some radius
in most expressions, we introduce the rescaled charges 
\begin{equation}
\wt Z_i \,=\, \frac{Z_i \,\ell_B}{a_1}
\end{equation}
Then, $X_1$ can be naturally expressed as a function 
of $\wt Z^1_\eff$ and of a dressed packing fraction 
\begin{equation}
\wt \eta \,=\, \frac{4 \pi}{3} \rho a_1^3,
\end{equation}
leading to
\begin{equation}
X_1 \,=\, \wt\eta \, \wt Z_\back .
\end{equation}
The dressed fraction $\wt\eta$ is connected to the packing fraction $\eta$ in the suspension
through 
\begin{equation}
\eta \,=\, \sum_i \frac{4 \pi}{3} \rho_i \, a_i^3 \, = \, \wt\eta \,\sum_i x_i \,\frac{a_i^3}{a_1^3}
.
\end{equation}

\begin{figure}[ht]
\begin{center}
\includegraphics[angle=0,width=8cm,clip]{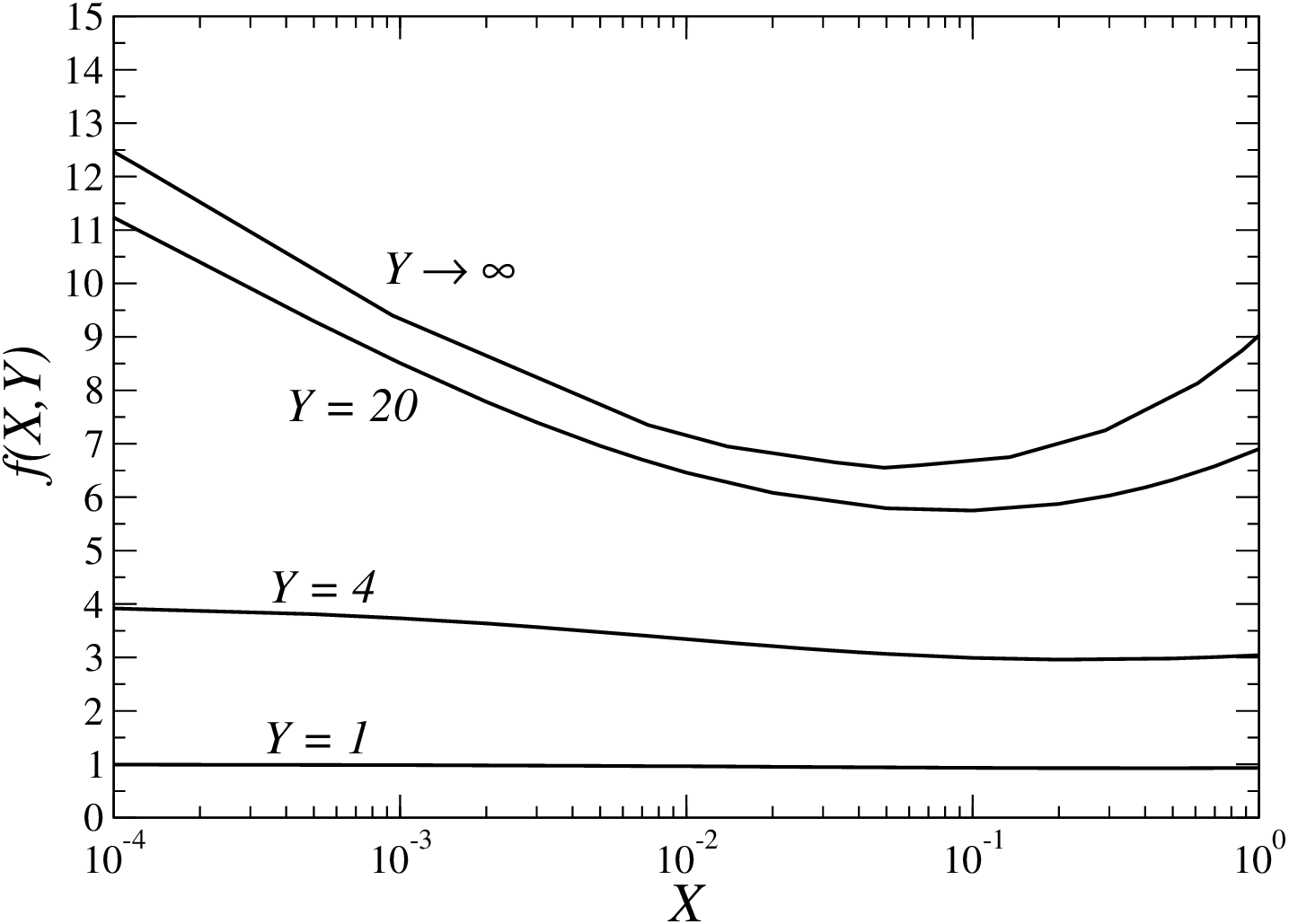}
\end{center}
\caption{Behavior of the effective charge $f(X,Y)$ as a function of screening,
as encoded in $X$. The quantity $Y$ denotes the bare charge of the macroion under study, so that the upper
curve, showing $f(X,\infty$) corresponds to the saturation value studied
in Ref. \cite{tba02}. Practically, $f(X,Y) \, a/\ell_B$ is the effective jellium charge of a sphere 
having radius $a$, bare charge $Y a/\ell_B$, at a packing fraction $X/f(X,Y)$
(mono-component case).
}
\label{fig:f}
\end{figure}

To summarize the previous discussion, the key equation to be solved within the jellium model 
is
\begin{equation}
\wt Z_\back \, \equiv \, \frac{Z_\back \,\ell_B}{a_1} \,=\, \sum_i x_i \frac{a_i}{a_1} \,
f\left( \wt\eta\, \wt Z_\back \,\frac{a_i^2}{a_1^2}, \wt Z^i_\bare \frac{a_1}{a_i}
\right).
\label{eq:summary_selfcons}
\end{equation}
Hence, once the physical parameters have been chosen (bare charges, compositions $x_i$, radii $a_i$ and packing fraction),
one needs to find the root $X^*$ of equation
\begin{equation}
\frac{X}{\wt\eta} \,=\, 
\sum_i x_i \frac{a_i}{a_1} \,
f\left( X \,\frac{a_i^2}{a_1^2}, \wt Z^i_\bare \frac{a_1}{a_i}
\right).
\label{eq:Ximport}
\end{equation}
from which the background (effective) charge follows: $\wt Z_\back = X^*/\wt\eta$.
Of course, the function $f(x,\wt Z_\bare^i a_1/a_i)$ should be computed before hand for
all species, but this task deals with a mono-component problem only.
In other words, $f(X,Y)$ is the effective charge of the potential $\phi$ obeying
\begin{equation}
\frac{d^2 \phi}{d \tilde r^2} + \frac{2}{\tilde r}  \frac{d \phi}{d \tilde r} \, = \, 
3 \, X (e^\phi-1),
\end{equation}
with boundary conditions
\begin{equation}
\phi \to 0 \hbox{ for } \tilde r\to \infty \quad \hbox{ and } \quad 
\frac{d \phi}{d\tilde r} = - Y \hbox{ at } \tilde r=1,
\end{equation}
meaning that for large $\tilde r$
\begin{equation}
\phi \sim f(X,Y) \,\,\frac{e^{-\tilde\kappa(\tilde r-1)}}{(1+\tilde\kappa) \, \tilde r} \quad \hbox{ with }\quad
\tilde \kappa^2 \,=\, 3\, X .
\end{equation}
It is thus straightforward to obtain $f$, following for instance the method presented in
the appendix of Ref.
\cite{tbavg03,rque50}. 
Typical results are shown in Fig. \ref{fig:f}.
When $Y$ is small, charge renormalization effects disappear, so that 
$f(X,Y)=Y$, irrespective of $X$. In the limit of small bare charges,
the background charge thus takes a simple form: 
$Z_\back = \sum_i x_i Z_\bare^i$.
On the other hand, upon increasing
the bare charge through $Y$, the effective charge also increases,
with always $f<Y$ \cite{rque10}. The saturation upper curve is reached
for large $Y$.

It appears at this point that the packing fraction (either the real one, $\eta$, or its dressed
counterpart $\wt\eta$), only enters the self-consistency condition 
on the left hand-side of Eq. (\ref{eq:Ximport}). As a consequence, our 
method allows to treat very simply the effect of packing fraction, since the more time consuming part of the calculation
is that of the right hand-side of Eq. (\ref{eq:Ximport}). This is an important advantage
over previous proposals, be it the technique presented
in \cite{ptl07}, or subsequent improvements \cite{castaneda1,castaneda2}. 

\begin{figure}
\begin{center}
\includegraphics[angle=0,width=8cm,clip]{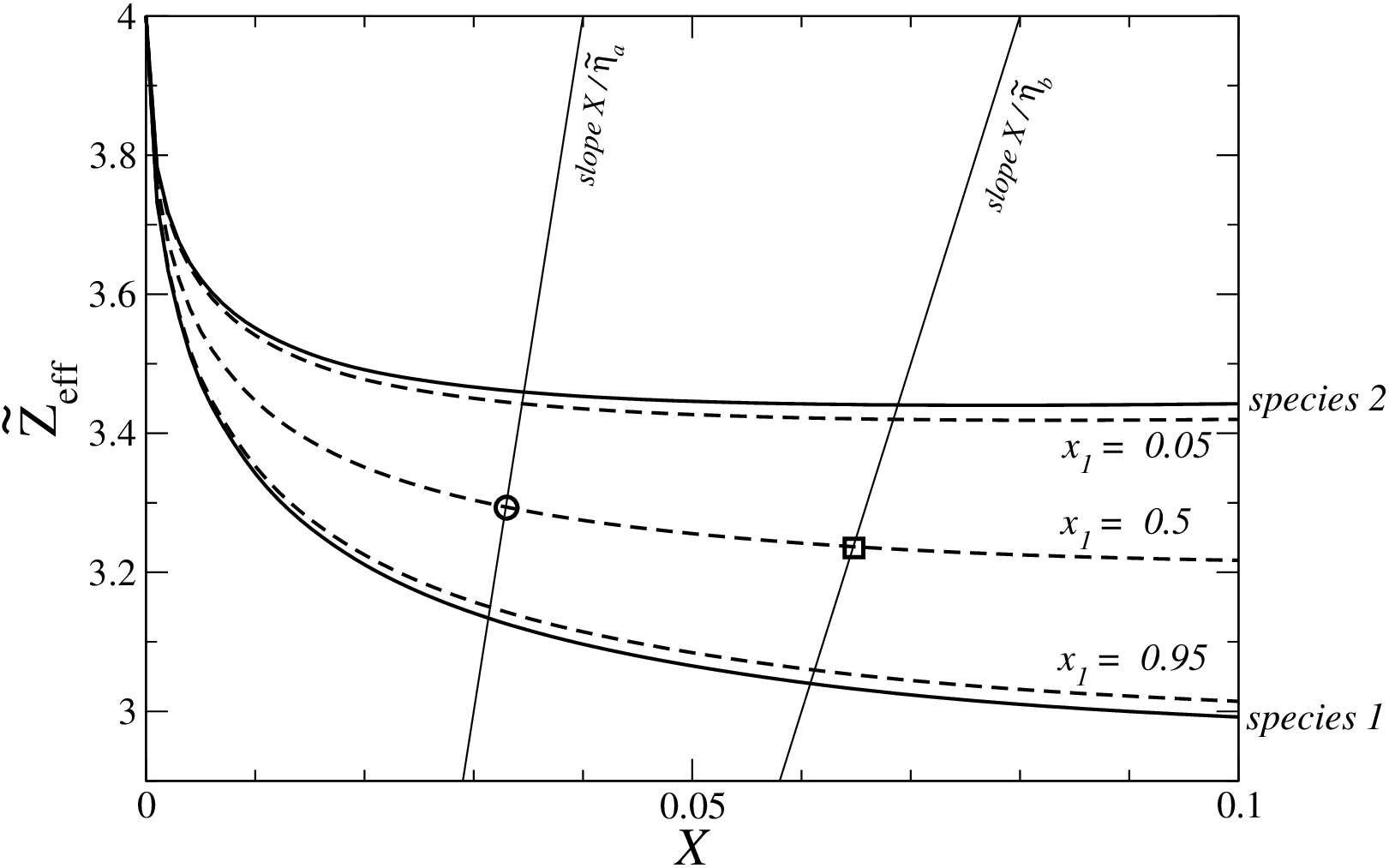}
\end{center}
\caption{Illustration of the method employed to find the solution of Eq. (\ref{eq:Ximport}),
for $\wt Z_\bare^1=\wt Z_\bare^2= 4$, $a_2=2\,a_1$, and $\wt\eta=10^{-2}$.
The continuous curves show the effective charge 
$f(X,4)$ (lower curve, indexed `species 1') and $2f(4X,2)$ (upper curve, indexed `species 2').
Depending on the mixture composition, the weighted average of both with weights
$x_1$ and $x_2=1-x_1$ are shown with the dashed lines. These are the master curves, corresponding to the right hand-side
of Eq. (\ref{eq:Ximport}), to be considered for all
possible $\wt\eta$. The linear curves show $X/\wt\eta$ for two values of the dressed packing fraction
($\wt\eta_a=10^{-2}$ and $\wt\eta_b=2\times 10^{-2}$). For an equimolar mixture ($x_1=1/2$), the effective background charge is shown,
by the circle (case $\wt\eta=10^{-2}$) and by the square (case $\wt\eta=2\times 10^{-2}$).}
\label{fig:method}
\end{figure}

\begin{figure}
\begin{center}
\includegraphics[angle=0,width=7cm,clip]{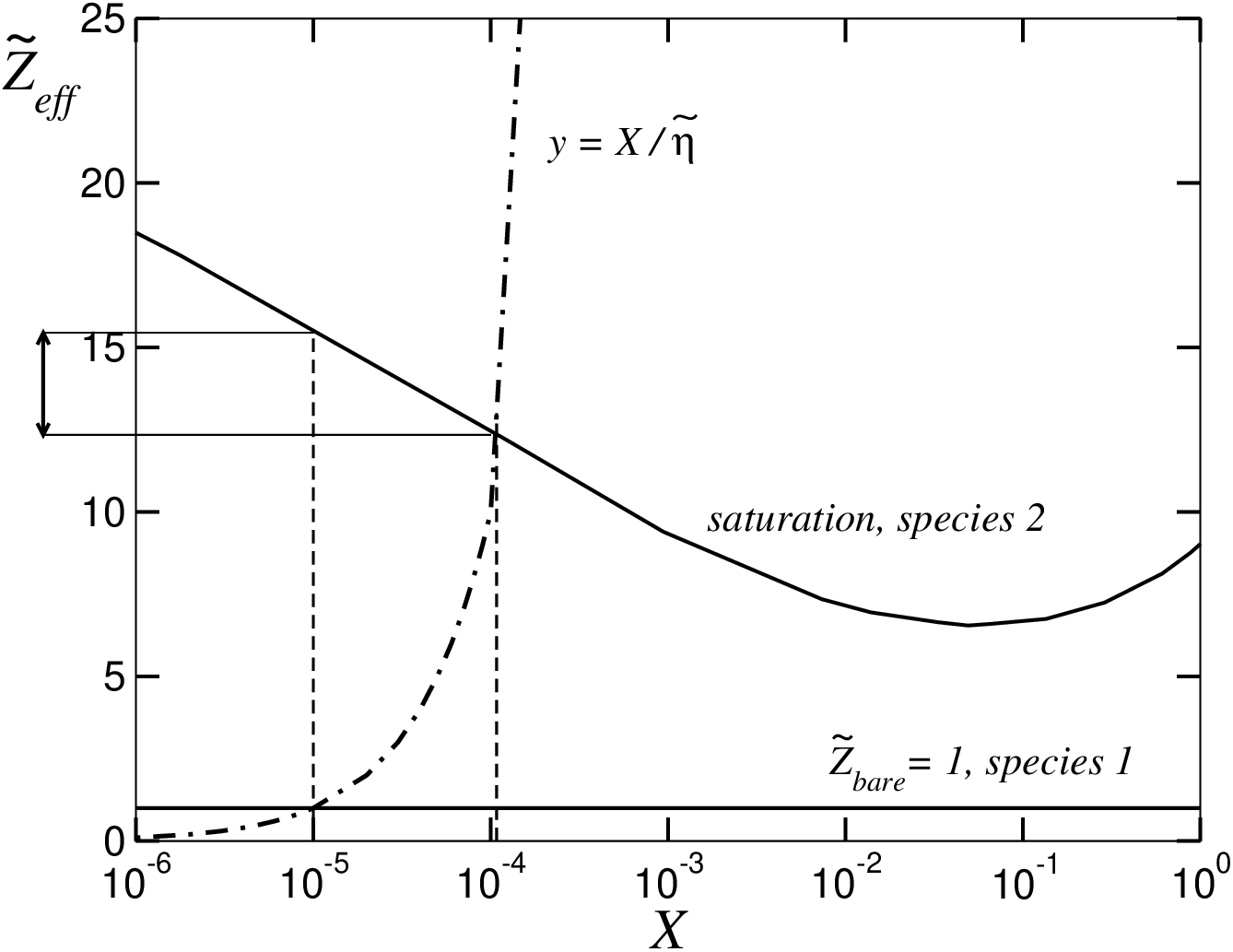}
\end{center}
\caption{Like-size binary mixture of a weakly charged species with $\wt Z_\bare^1=1$ and
a strongly charged species (limit $\wt Z_\bare^2 \to \infty$). The packing fraction is $\eta=\wt\eta=10^{-5}$.
When changing the mixture composition, the allowed range for $\wt Z_\eff^2$ is displayed
by the vertical double arrow on the l.h.s.).
}
\label{fig:1:sat}
\end{figure}

\begin{figure}
\begin{center}
\includegraphics[angle=0,width=7cm,clip]{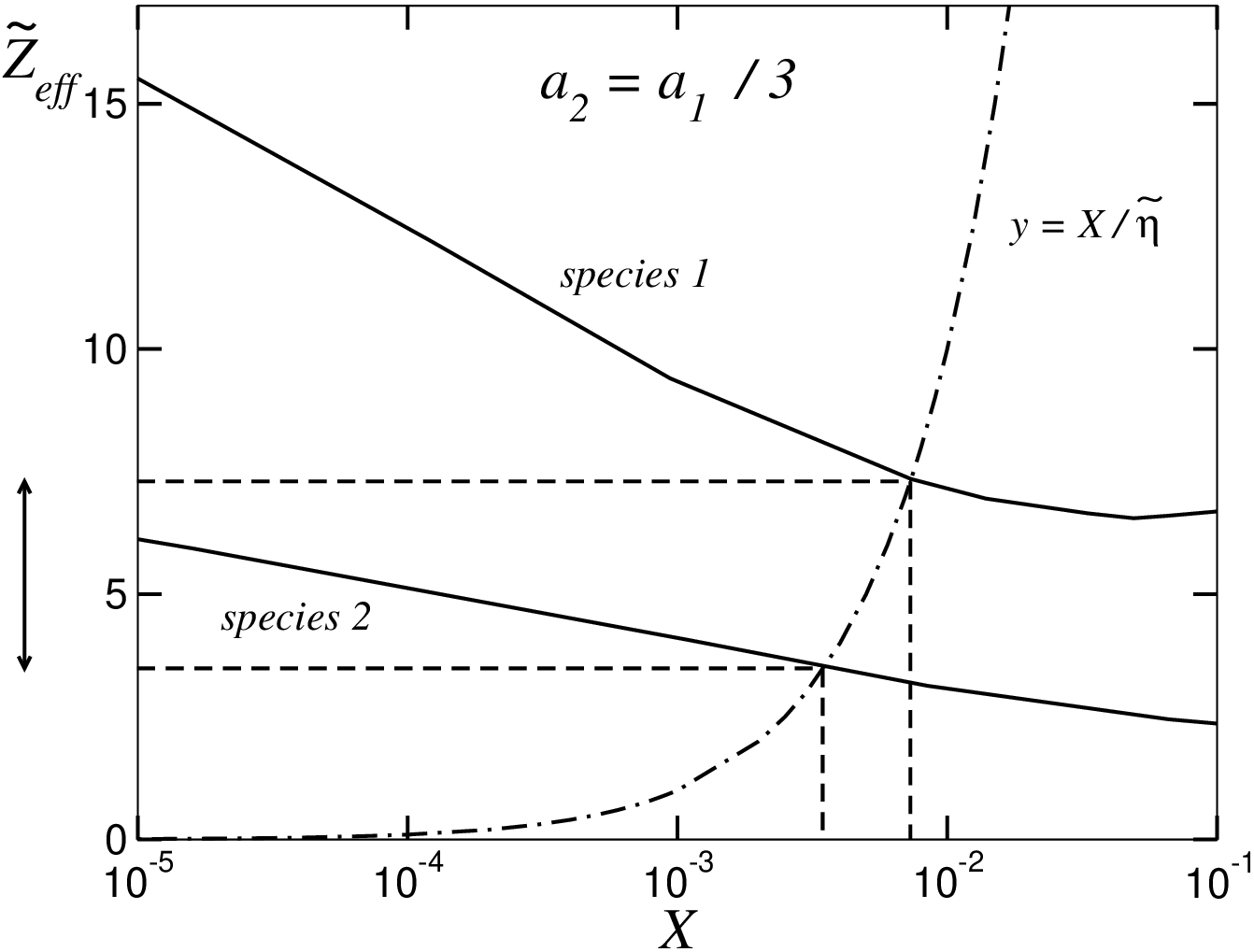}
\end{center}
\caption{Binary case. Log-linear plot. Here $\wt Z_\bare^1=\wt Z_\bare^2$ are both saturated (divergent), $\wt\eta=10^{-3}$, $a_2/a_1=1/3$. 
The weighted average (\ref{eq:summary_selfcons}) lies in between the two thick curves upon changing the composition $x_1$
from 0 (in which case it corresponds to the `species 2' bottom curve) to 1 (in which case it coincides to the `species 1' upper curve). As a consequence,
the values $X$ can take lie between the two vertical dashed lines, from which the allowed range 
for $\wt Z_\back$ can be read on the $y$-axis, and falls in between the two horizontal dashed lines.
As in Fig. \ref{fig:1:sat}, the allowed range is thus shown by the vertical arrow. 
}
\label{fig:ratio0.333}
\end{figure}

\begin{figure}
\begin{center}
\includegraphics[angle=0,width=8cm,clip]{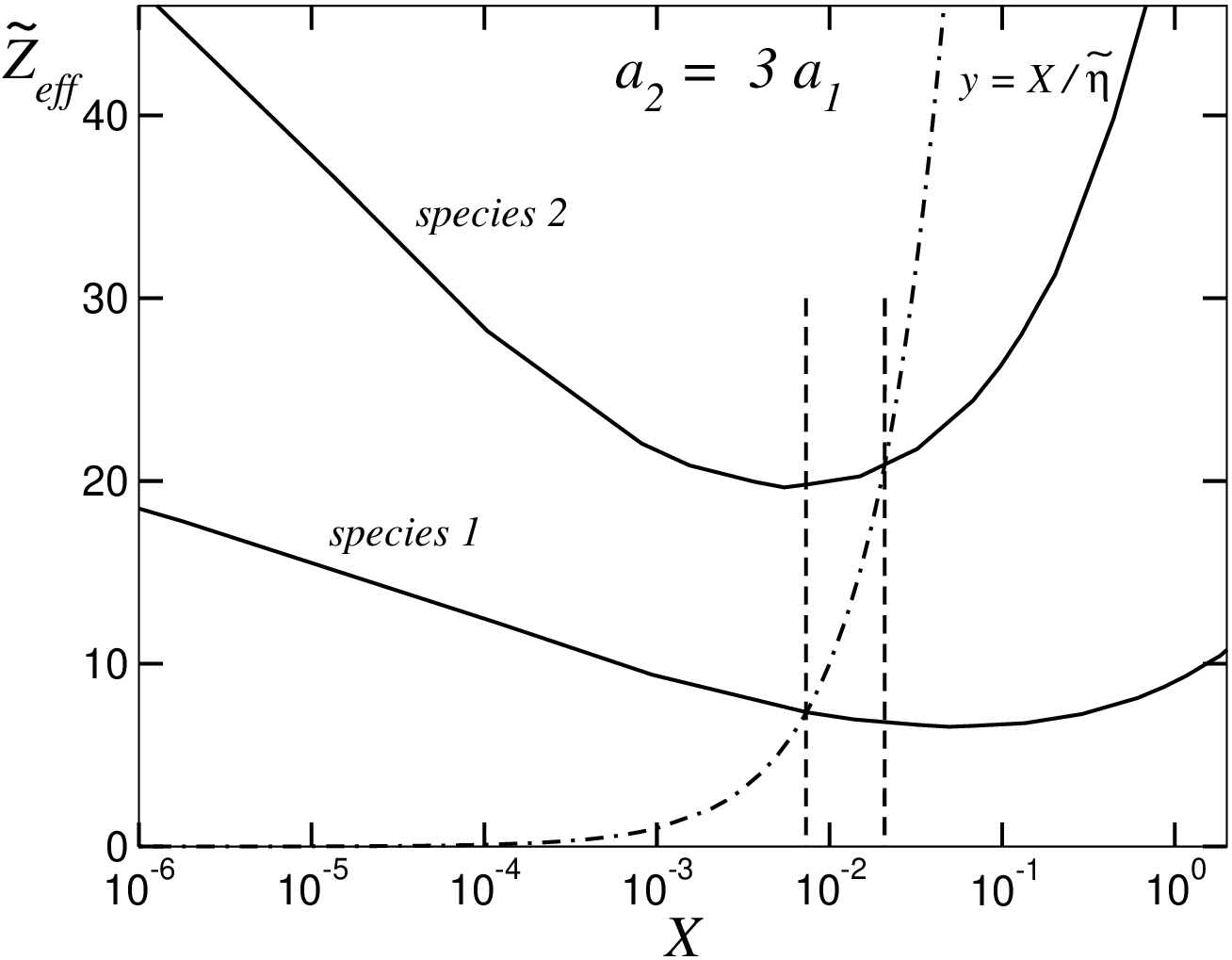}
\end{center}
\caption{Same as Fig. \ref{fig:ratio0.333} but for $a_2/a_1=3$.}
\label{fig:ratio3}
\end{figure}

For concreteness, the explicit solution of a binary colloidal problem is constructed
in Fig. \ref{fig:method} with relatively weakly charged macroions: both have the same charge $\wt Z_\bare^1=\wt Z_\bare^2= 4$,
but they differ in size: $a_2/a_1=2$. The pristine effective charges $f(X,4)$ and $f(X,2)$ should be known, from which one constructs 
the weighted average appearing in the r.h.s of Eq. (\ref{eq:summary_selfcons}) is calculated. Depending on the mixture composition,
this leads to the dashed curves: from bottom to top are a species 1-rich, an equimolar and a species 2-rich mixture.
The procedure closes, after the choice of density through $\wt\eta$, by searching for the intersection with
the line $X/\wt\eta$. With $x_1=1/2$, we thereby get the background charge $\wt Z_\eff = 3.29$ at $\wt\eta=10^{-2}$,
and $\wt Z_\eff = 3.23$ at $\wt\eta=2\times 10^{-2}$. The graphical construct proposed allows
to anticipate the dependence of effective charges on mixture composition, see Fig. \ref{fig:1:sat}
which corresponds to a bi-disperse solution with $a_1=a_2$ but unequal bare charges.
It can be expected that increasing $\eta$, a regime will be reached in the vicinity of the 
species 2 curve minimum, where the 
corresponding range for the variations of $Z_\sat^2$ with composition
will vanish. This will be confirmed in Section \ref{sec:results}.
Turning to the effect of binary mixture composition on background charge in the case of
unequal colloidal sizes, Figures \ref{fig:ratio0.333} and \ref{fig:ratio3}
address large bare charges (saturated limit) and show by vertical dashed lines
how $X$ is affected by going from $x_1=1$ to $x_1=0$. 
Once $X$ (or more precisely, the root $X^*$) is known, the background charge
follows from $\wt Z_\back = X/\wt\eta$. These two figures are for $a_2/a_1 = 1/3 $ and 3.
Of course, the $1\leftrightarrow 2$ labeling of species is immaterial in the case $x_1=x_2=1/2$, so that at a given density 
$\rho$, the solutions of the two problems should coincide. This is not the case 
in Figs. \ref{fig:ratio0.333} and \ref{fig:ratio3} since $\wt \eta$ is common to both,
meaning that they correspond to different densities $\rho$.

\subsection{The general case}
\label{ssec:salt}

\begin{figure}
\begin{center}
\includegraphics[angle=0,width=8cm,clip]{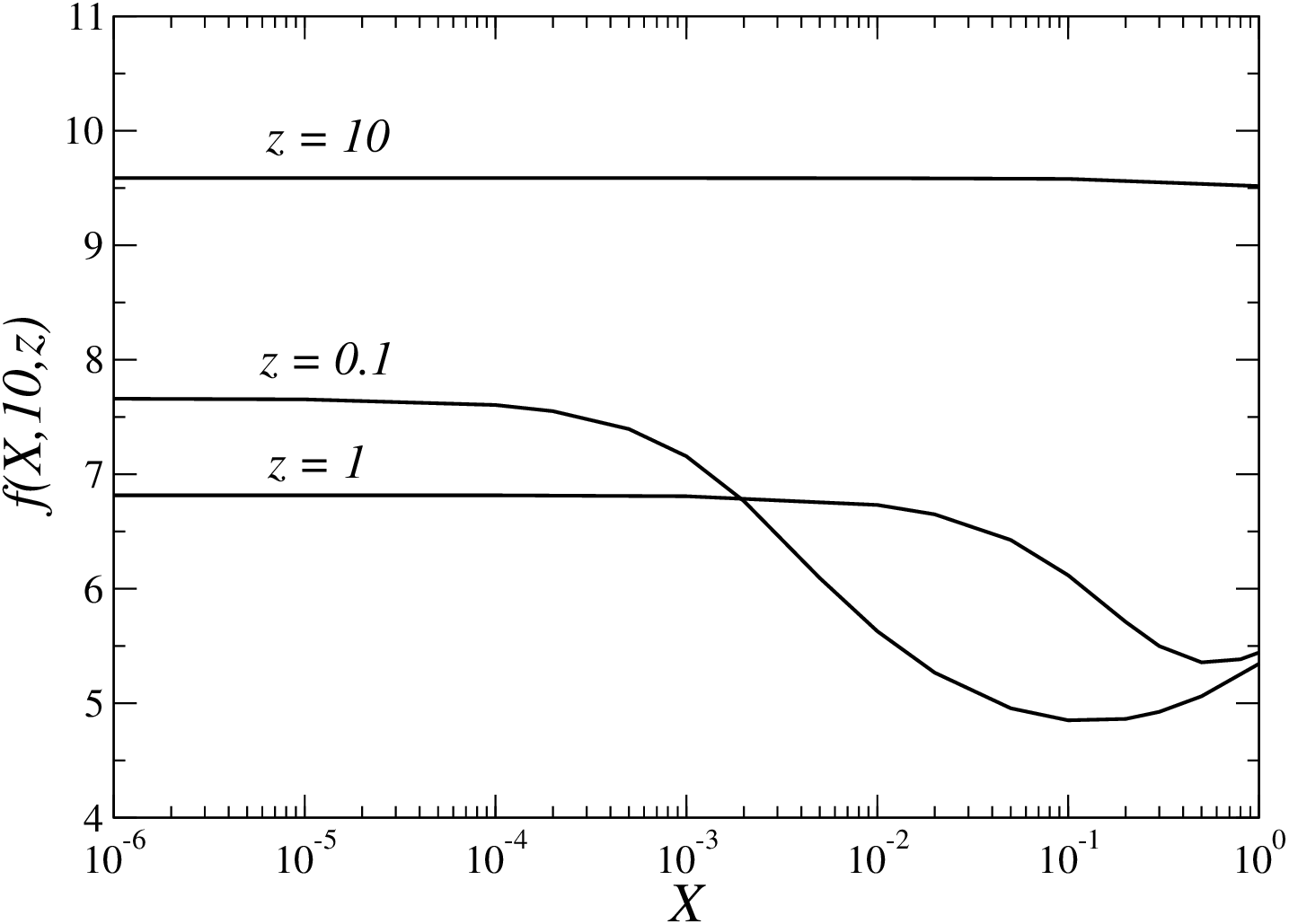}
\end{center}
\caption{Effect of salt on the screening function $f_\salt$ appearing 
in Eq. (\ref{eq:fdef}). Here, the reduced charge in chosen equal to 10,
and we show $f_\salt(X,10,z)$ as a function of $X$ (the jellium background dimensionless 
charge), for different salinities
$z$.}
\label{fig:f_salt}
\end{figure}

So far, the discussion focused on the deionized limit. In case salt is present,
for instance when the system is in osmotic equilibrium with a salt reservoir of density
$c_s$,
Eq. \eqref{eq:Poisson_nosalt} becomes
\begin{equation}
\nabla^2 \phi_i \,=\, 4 \pi \ell_B \,\left[2c_s \sinh\phi_i -\rho Z_{\text back}
\right],
\label{eq:Poisson_salt_0}
\end{equation}
with the boundary conditions:
\begin{equation}
2c_s \sinh\phi_i -\rho Z_{\text back} \to 0  \hbox{ for } r  \to \infty \quad \hbox{ and } \quad 
\frac{d \phi_i}{dr} = - Z^i_{\text bare} \ell_B\frac{1}{a_i^2} \hbox{ at } r=a_i.
\label{eq:bcsalt}
\end{equation}
The first equation stems from electroneutrality and defines the potential at infinity, often 
referred to as the Donnan potential.
The second results from Gauss' theorem.
Defining the inverse squared Debye length in the reservoir as 
$\kappa_\res^2 = 8 \pi \ell_B c_s$, we arrive at 
\begin{equation}
\nabla^2 \phi_i \,=\, \kappa_\res^2 \sinh\phi_i - 4 \pi \ell_B\,\rho Z_{\text back},
\label{eq:Poisson_salt}
\end{equation}
and we can proceed along very similar lines as in Section \ref{ssec:nosalt}.
We have assumed here the salt to be monovalent, for simplicity.
Generalization to mixed-valency salts is straightforward.
Expressing the colloids' effective charges requires the introduction of a
generalization of function $f$, which we denote $f_\salt$, so that
\begin{equation}
\frac{Z^i_{\eff} \,\ell_B}{a_i} \,=\, f_\salt\left(X_i, \frac{Z^i_{\bare}\,\ell_B}{a_i}, 
\kappa_\res a_i\right),
\label{eq:fdef}
\end{equation}
keeping the same notation for $X_i$.
Of course, one has $f(X,Y) = f_\salt(X,Y,0)$.
The self-consistency condition becomes 
\begin{equation}
Z_\back  \,=\, \sum_i x_i \, Z^i_{\text{eff}} \,=\,
\frac{1}{\ell_B} \sum_i x_i\, a_i\, f_\salt\left(4 \pi \ell_B \rho Z_{\back} a_i^2/3,\frac{Z^i_{\bare}\,\ell_B}{a_i},\kappa_\res a_i\right).
\label{eq:selfcons_salt}
\end{equation}
Again, the functions $f_\salt$, which are those of a single component problem, can be
computed as such \cite{tbavg03}, and subsequently used to describe an 
arbitrary mixture. Typical results are shown in Fig. \ref{fig:f_salt},
for a colloidal bare charge that is neither small nor large,
meaning that it is of order $10 \,a/\ell_B$.

From the very form of Eq. \eqref{eq:Poisson_salt}, it appears that the long distance 
potential $\phi_i$ is of the standard form 
\begin{equation}
\phi_i -\phi_i(\infty) \, \sim \, \frac{Z_\eff^i \, \ell_B}{(1+\kappa a_i)} \,\frac{e^{-\kappa (r-a_i)}}{r}
\end{equation}
an expression which can be viewed as defining the effective charge $Z_\eff^i$,
and which involves the effective screening length $\kappa^{-1}$ given by
\begin{equation}
\kappa^2 = \kappa_\res^2 \, \cosh [\phi_i(\infty)].
\end{equation}
This quantity can be re-expressed as 
\begin{equation}
\kappa^4 \, = \, \kappa_\res^4 \,+\, \left(4\pi\ell_B\,\rho\,Z_\back   \right)^2.
\end{equation}

It is worth emphasizing here that a {\it bona fide} feature of jellium-like models
is that the osmotic pressure takes a particularly simple form, and is directly
connected to the effective charges \cite{tl04,ptl07,castaneda2}:
\begin{equation}
\beta P \,=\, \rho \,+\, 2 c_s \cosh[\phi_i(\infty)] -2 c_s \,= \,
\rho \,+\, \sqrt{(2 c_s)^2 + (\rho Z_\back)^2} -2 c_s .
\end{equation}
It is the excess pressure with respect to the salt reservoir, including the colloidal
contribution, taken ideal for simplicity. For salt-free systems,
it takes the form $\beta P = \rho + Z_\back \rho$, which is usually close
to $Z_\back \rho$.

\subsection{Comparison with previous approaches}
Before discussing the physical results, it seems opportune to put the method 
described above in the context of those used so far. For the sake of the discussion,
we assume that the salt content is fixed, and we wish to identify the number
of independent parameters that have to be (essentially continuously) varied before the full solution
is reached. This allows for a definition of the 'dimensionality' of the method,
a measure of user-friendliness.

We start by the mono-component case, and consider that the goal is to obtain 
a curve $Z_\eff$ as a function of $Z_\bare$, parameterized by $\eta$.
The original method used in Refs \cite{tl04,ptl07} is brute force:
for each $\eta$, $Z_\bare$ and $Z_\back$,  Eq. (\ref{eq:Poisson_salt}) is solved
by a shooting method, to obtain the desired value of $Z_\bare$: this is a
procedure of dimension 1 \cite{rque51}. Then $Z_\back$ should be changed,
to find in which case the background and effective charges coincide.
In that respect, the resolution is of dimension 2 for each $\eta$ and $Z_\bare$, it is thus
of dimension 4 overall. Casta\~neda-Priego and collaborators \cite{castaneda1,castaneda2} have found an interesting
reformulation, in which self-consistency is automatically enforced by imposing {\it a priori}
$Z_\back=Z_\eff$, and computing the corresponding $Z_\bare$ in one step only. This is achieved by 
constraining the far-field. For each 
$\eta$, the method is of dimension 1 ($Z_\eff$ has to be changed). 
Hence, the overall dimension is two, which is an improvement.
Finally, with the method presented here, 
a unique function $f$ of two parameters encodes the relevant information, and 
the approach also is of dimension 2.

The `degeneracy' between the latter two procedures is lifted when considering 
mixtures. Following Ref. \cite{castaneda1,castaneda2}, the effective charges have to be chosen 
{\it a priori}, and the bare charges follow. However, a physical 
problem is in practice formulated in terms of bare charges. This subtlety is immaterial
for mono-component systems: the functions $Z_\eff(Z_\bare)$ and $Z_\bare(Z_\eff)$ convey the
same information, and are simply connected. This is no longer the case for mixtures, where the functions
$Z_\eff^i(Z_\bare^1,Z_\bare^2\ldots)$ and $Z_\bare^i(Z_\eff^1,Z_\eff^2\ldots)$ are not simply related.
Deriving the second from the first requires a shooting task that appears quite impractical.
Additionally, there is no guarantee that 
the {\it a priori} choices of effective charges are not unphysical, with for instance values
above the saturation limit. This is the case for instance in Fig. 5 of Ref.
\cite{castaneda2}, for low salt content \cite{rque60}.
Our alternative treatment is free of these shortcomings.

\section{Results}
\label{sec:results}

\subsection{General features of effective charges}

\begin{figure}
\begin{minipage}{0.48\linewidth}
\begin{center}
\includegraphics[angle=0,width=8cm]{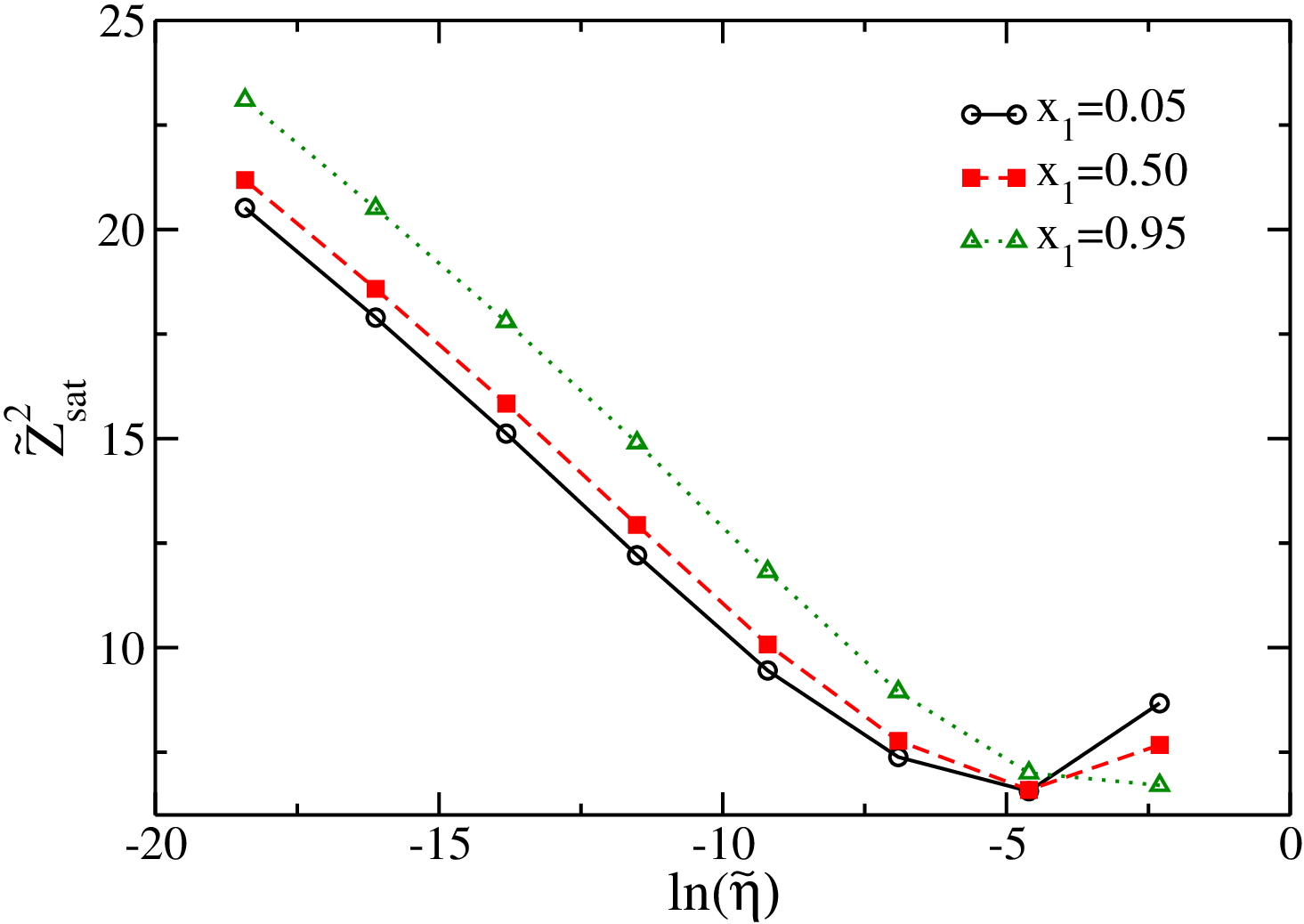}
\end{center}
\end{minipage}
\begin{minipage}{0.48\linewidth}
\begin{center}
\includegraphics[angle=0,width=8cm]{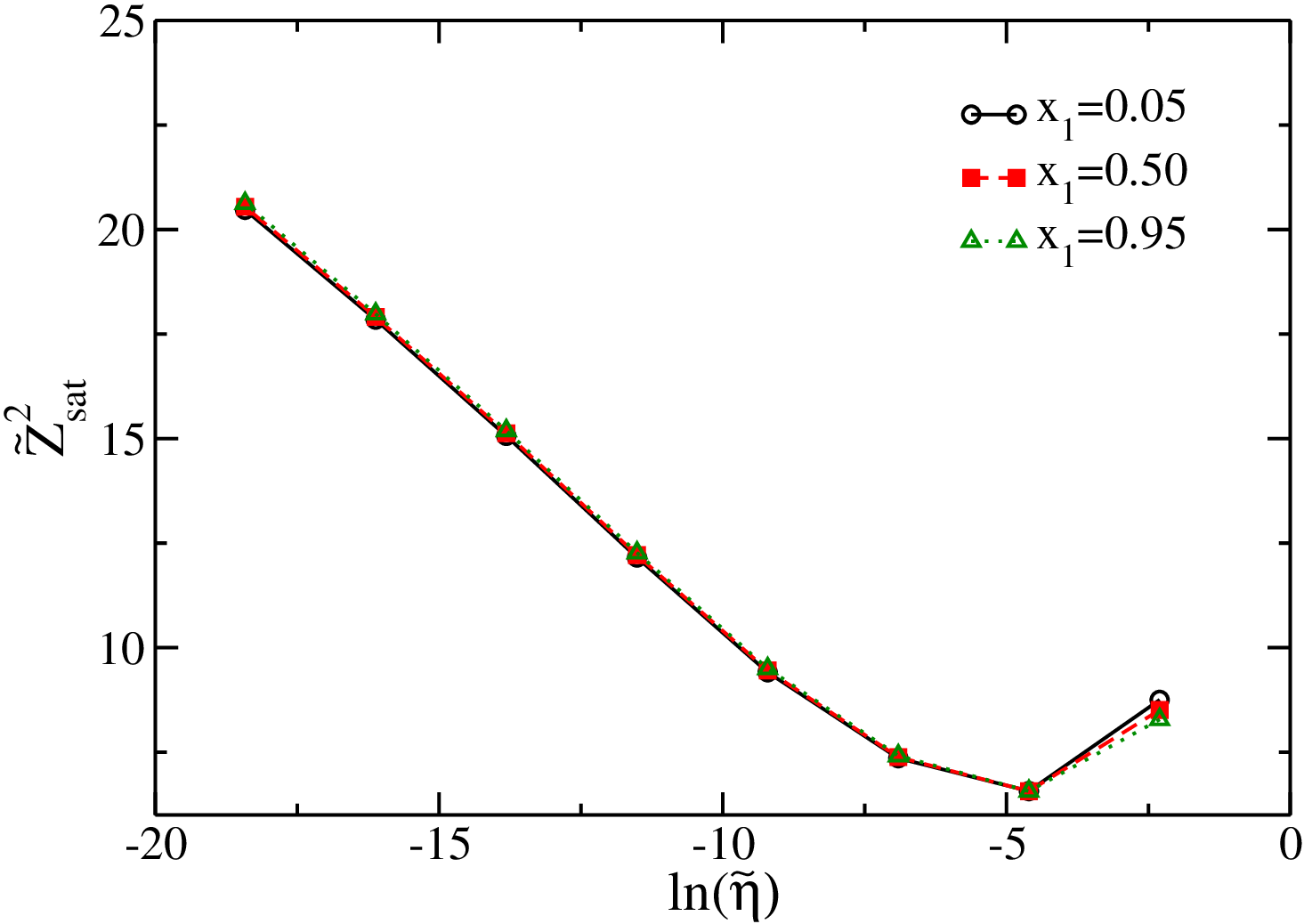}
\end{center}
\end{minipage}
\caption{Saturation charge, $\widetilde{Z}_{\text{sat}}^2$, as a function of the total density of colloids in the no salt case. 
The dependence on $x_1$ is not very strong. In these case  $\widetilde{Z}_{\text{bare}}^1=1$ (left) and $\widetilde{Z}_{\text{bare}}^1=20$ (right).
}\label{saturacionbis}
\end{figure}

In this section we focus on the behavior of the saturation charge. In \cite{tl04}, it has been found that the saturation value for the charge when the concentration 
was small ($\widetilde \eta<10^{-5}$) was given by
\begin{equation}
\label{ec:sat}
Z_{\text{\text{sat}}}\simeq\frac{a}{\ell_B}[\delta-\gamma \log(\widetilde \eta)],
\end{equation}
where $\gamma\simeq 1$ and $\delta\simeq 2$.  In Fig. \ref{saturacionbis} the saturation value $\widetilde{Z}_{\text{sat}}^2$ has been plotted as a function of the density $\widetilde \eta$ 
for the no salt case, for  $\widetilde{Z}_{\text{bare}}^1=1$ (left) and $\widetilde{Z}_{\text{bare}}^1=20$ and 3 values of the composition $x_1$. As we can see, for small 
values of $\widetilde \eta$, equation (\ref{ec:sat}) holds, with 
different values for $\delta$ and $\gamma$, that depends slightly on $x_1$. In Fig. \ref{saturacionbis} (right) we reobtain the monodisperse case because both species are of 
the same size and the bare charges are large enough to be  
in the saturation limit.

\begin{figure}
\vspace*{1.5cm}
\begin{center}
\includegraphics[angle=0,width=8cm]{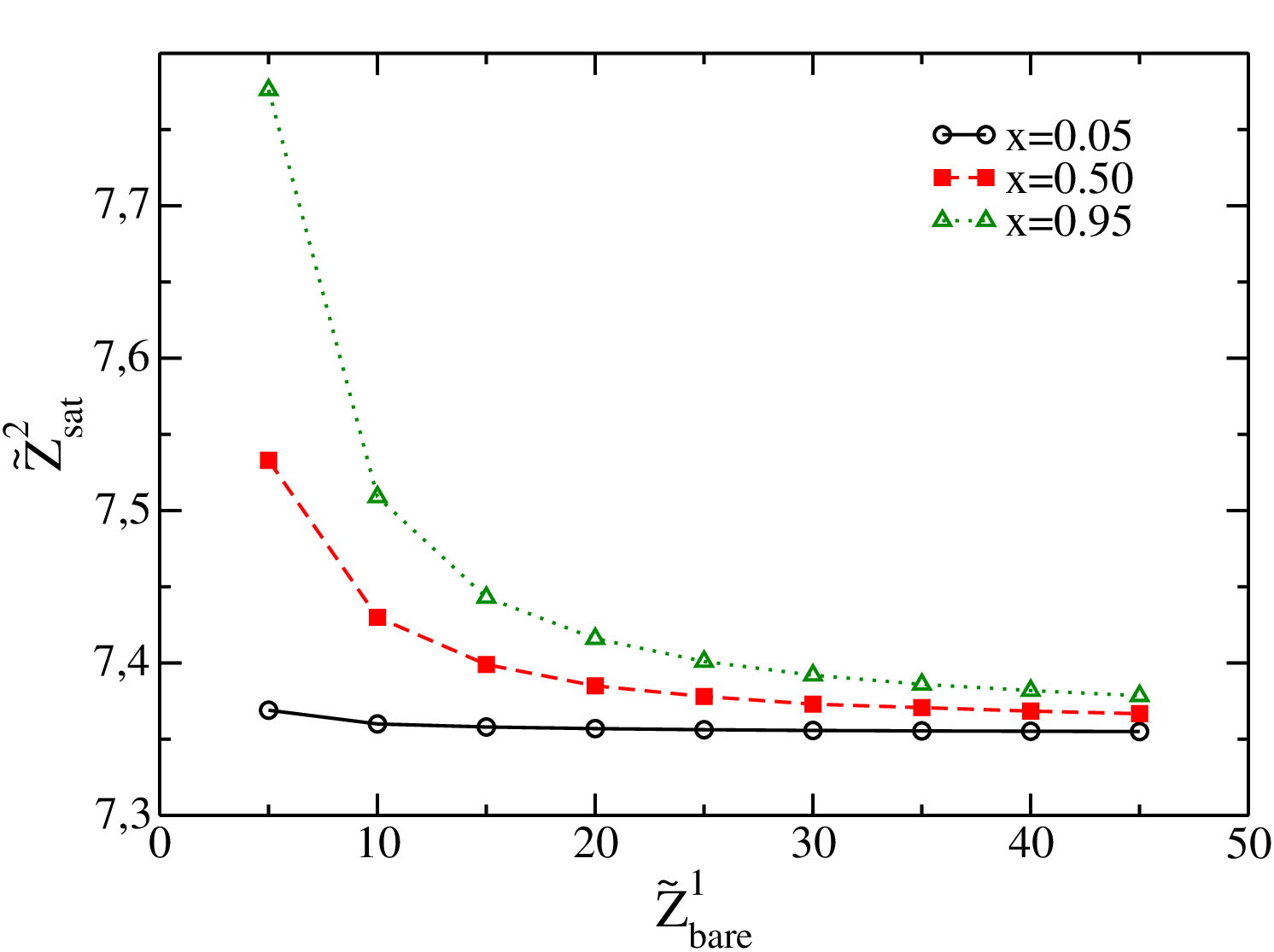}
\end{center}
\caption{Saturation charge in the no salt case as a function of $\widetilde{Z}_{\text{bare}}^1$ for a value of the total fraction $\widetilde \eta=10^{-3}$, and with $a_1=a_2$
(so that $\eta=10^{-3}$ as well).
}\label{saturacion2}
\vspace*{0.25cm}
\end{figure}

\begin{figure}
\begin{minipage}{0.48\linewidth}
\begin{center}
\includegraphics[angle=0,width=8cm]{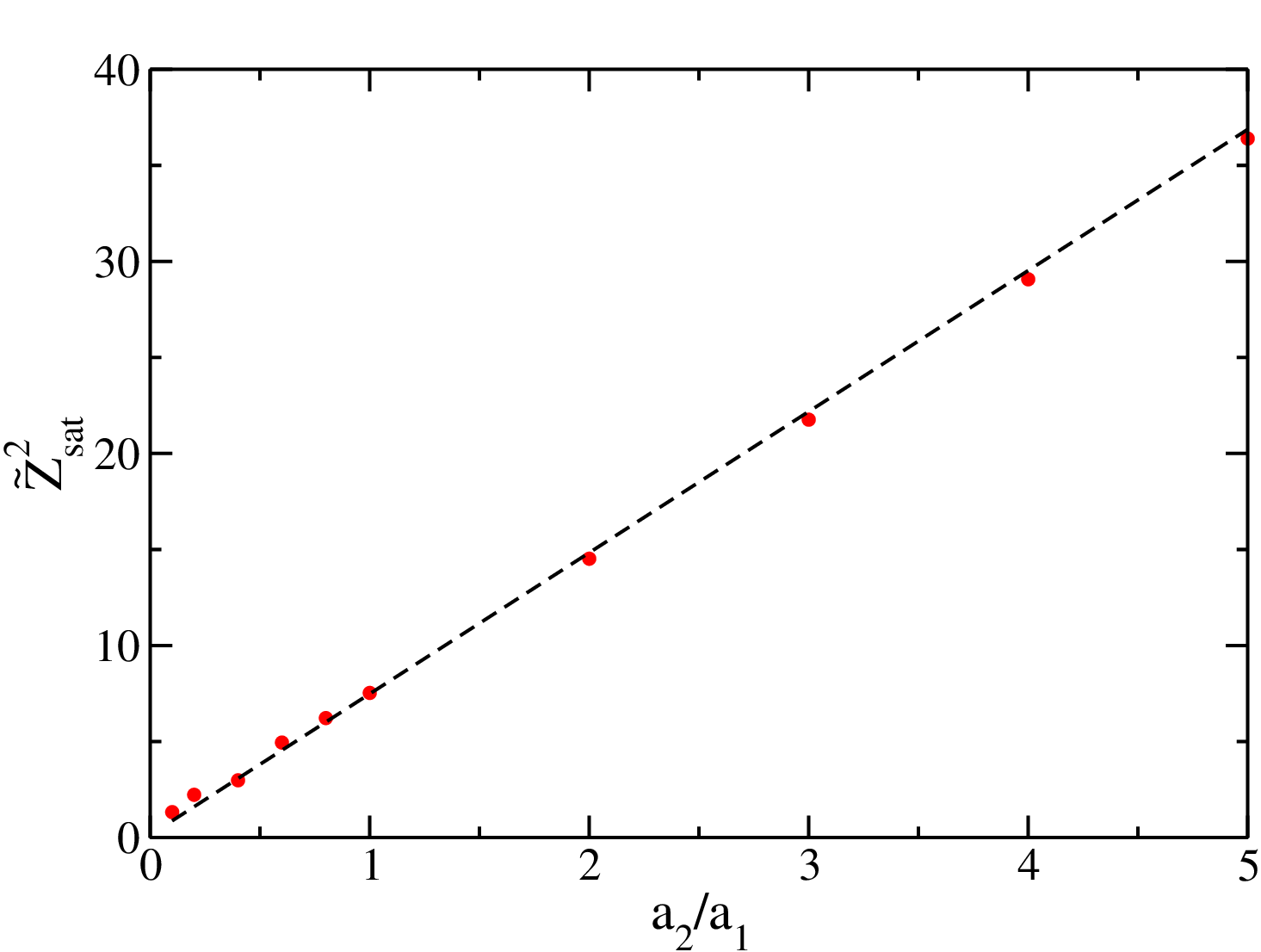}
\end{center}
\end{minipage}
\begin{minipage}{0.48\linewidth}
\begin{center}
\includegraphics[angle=0,width=8cm]{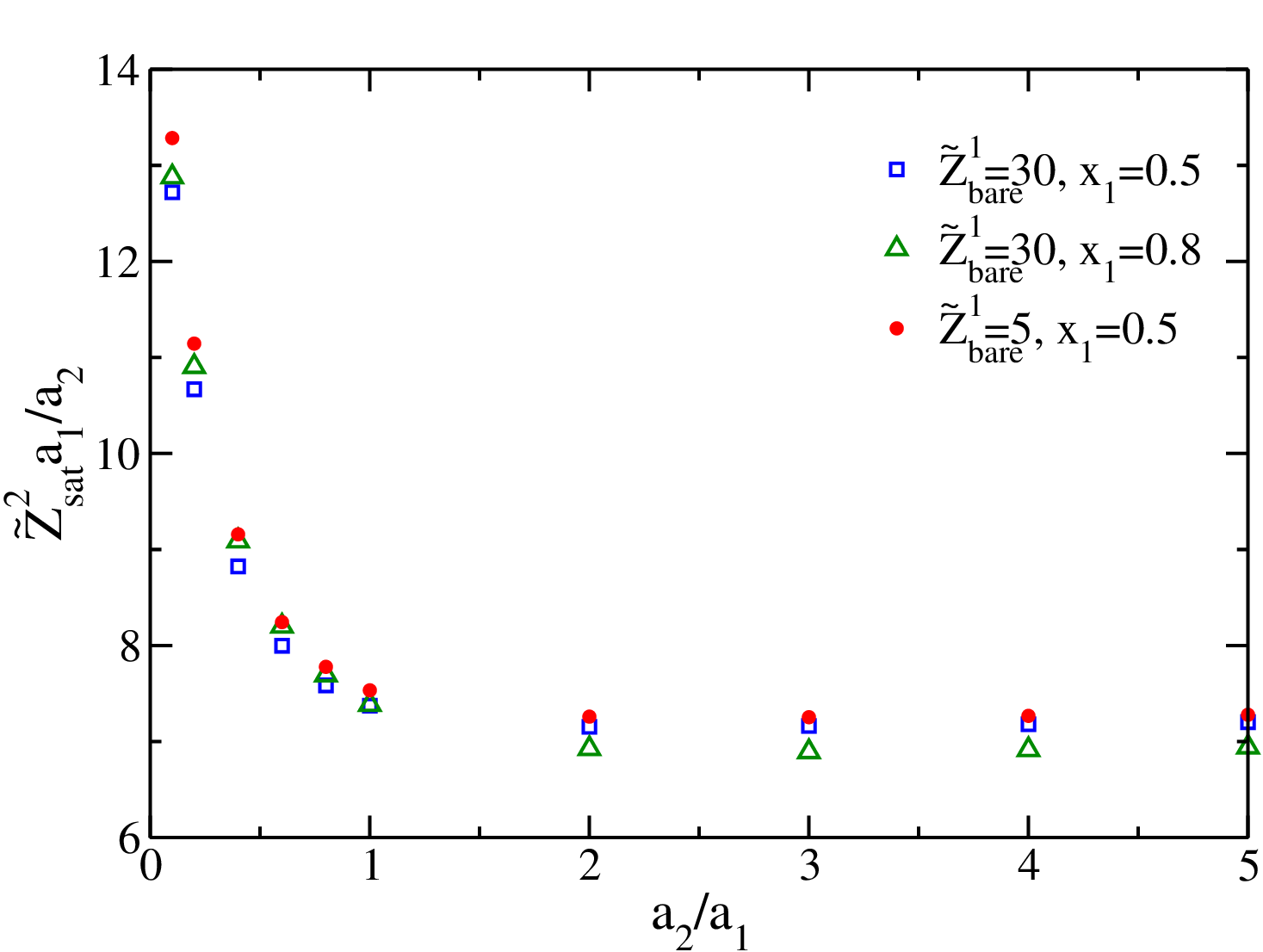}
\end{center}
\end{minipage}
\caption{(Left) Saturation charge $\widetilde{Z}_{\text{sat}}^2$ as a function of the radius ratio $a_2/a_1$, for a system with $\widetilde \eta=10^{-3}$, $\widetilde{Z}_{\text{bare}}^1=5$ and $x_1=0.5$. (Right) $\widetilde{Z}_{\text{sat}}^2 a_1/a_2$ for a system with $\widetilde \eta=10^{-3}$ and different values of  
$\widetilde{Z}_{\text{bare}}^1$ and $x_1$.
}\label{saturacionfuncionb}
\end{figure}

In Fig. \ref{saturacion2}, the saturation value has been plotted as a function of  $\widetilde{Z}_{\text{bare}}^1$ for a density $\widetilde \eta=10^{-3}$ and 3 values of the concentration $x_1$. 
The dependence on $x_1$ decreases as the value of  $\widetilde{Z}_{\text{bare}}^1$ increases because we approach the saturation for species $1$.
We are now in a position to analyze the dependence of this property on the colloidal sizes asymmetry. To this aim, we have studied the variation of the saturation value of the charge  as we 
vary the size ratio. In Fig \ref{saturacionfuncionb}-left, we have plotted $\widetilde{Z}_{\text{sat}}^{2}$ as a function of $a_2/a_1$ for a system with $\widetilde \eta=10^{-3}$ and $x_1=0.5$.
It appears that the dependence is roughly linear on $a_2/a_1$. The dashed line is a linear fitting. However, on closer inspection, 
the situation is more complex ; see Fig \ref{saturacionfuncionb}-right
plotting $\widetilde{Z}_{\text{sat}}^{2}\,a_1/a_2$ for different values of $\widetilde{Z}_{\text{bare}}^1$ and $x_1$. It can be seen 
that for $a_2/a_1<1$, the  behavior of $\widetilde{Z}_{\text{sat}}^{2}$ is not linear in $a_2/a_1$. 
This behavior can be understood from the plot of $f(X,\infty)$ reported in Fig. \ref{fig:f}, which exhibits in its
left-most part (say for $X<10^{-2}$), the entropy dominated regime alluded to in the introduction
(decrease of the effective charge with an increase of concentration). 
Upon decreasing $a_2$ at fixed $a_1$, the relevant background parameter $X_2$ decreases as $a_2^2$,
and this leads, from Eq. (\ref{eq:blabla}), to an increase of $Z^2_\sat \ell_B /a_2$. On the other hand,
increasing $a_2$, one probes at some point the shallow minimum seen in Fig. \ref{fig:f},
where $f$ takes values around 7. This is compatible with Fig. \ref{saturacionfuncionb}-right,
and also means that $\widetilde{Z}_{\text{sat}}^2 = {Z}_{\text{sat}}^2 \ell_B/a_1$ scales like
$a_2/a_1$ (see Fig. \ref{saturacionfuncionb}-left).


\subsection{Osmotic pressure and comparison to Monte Carlo simulations}

One of the advantages of the jellium model is that, once the renormalized charges are known, the evaluation of the osmotic pressure is straightforward. 
However, a competing theory of equal simplicity does exist \cite{ttvr08}, where the standard Poisson-Boltzmann cell model \cite{Marcus,alex} has been generalized
for mixtures. For colloidal spheres, the radii of the cells can be different for each type of 
macro-ion. These radii are determined self-consistently for a given set of parameter,
from the solution of the nonlinear Poisson-Boltzmann equation with appropriate boundary conditions \cite{ttvr08}.

In this section, 
we compare the results from both methods, with those of Monte Carlo (MC) simulations of bidisperse systems of spherical 
charged colloids. Explicit counter-ions are considered, without added salt. The simulations, which treat exactly Coulombic forces, have been performed in the NVT ensemble
with periodic boundary conditions.
In order to take into account the long range electrostatic interactions with the images of the system, Ewald summations were used \cite{Frenkel,thesisCarlos}.  
The number of colloidal particles of each type is $N_1=N_2=40$, confined in a simulation box of side length $L$. 
The number of monovalent counterions, $N_{ion}$, was set in each case so that charge neutrality was obtained.\\ 

The pressure of the system was computed using the virial theorem
\begin{equation}
  \beta P=\rho+\beta\left<W\right>,
\end{equation}
where $\rho$ is the particle number density, $\beta=(k_BT)^{-1}$ and $W$ is the virial function
\begin{equation}
  W=-\frac{1}{3V}\sum_{i=1}^N\bm{r}_i\cdot\nabla_iU
\end{equation}
for a system with particles at positions $\bm{r}_i$ interacting between themselves with a pair potential $U$ which is the sum of the long range Coulomb potential, using the known Ewald expressions \cite{ewald,leeuw,allen} with the minimum image convention, and a short range hard core potential. 

In order to compute $\left<W\right>$ for the hard core part of the potential we use \cite{perram}
\begin{equation}
  \beta\left<W\right>=\frac{1}{3V}\left<\sum_{i=1}^{N-1}\sum_{j>i}^N2F(\bm{r}_{ij})\ \delta(F(\bm{r}_{ij})-1)\right>,
\end{equation}  
where $F(\bm{r}_{ij})$ is an overlap function. In the case of spherical particles the overlap function has a simple form and the virial expression for the hard core interaction is
\begin{equation}
  \beta\left<W\right>=\frac{1}{3V}\left<\sum_{i=1}^{N-1}\sum_{j>i}^N\frac{r_{ij}^2}{\sigma_{ij}}\ \delta\hspace{-0.1cm}\left(r_{ij}-\sigma_{ij}\right)\right>,
\end{equation} 
in which $\sigma_{ij}=(\sigma_i+\sigma_j)/2$ and $\sigma_i$ is the diameter of particle $i$.\\

In all the simulations, the radius of the first colloidal species ($a_1$) was kept constant, and used to normalize the distances. 
The radius of the ions was set to $a_{\text{micro}}= 10^{-3}a_1$. 
The volume of the simulation box and the Bjerrum length were also kept constant at $(L/a_1)^3=33540.8$ and $\ell_B/a_1=0.32$ respectively. 
The systems were equilibrated for $3\times10^5$ MC steps before averaging and then the averages were carried out for 
$3\times10^5\sim 8\times10^5$ MC steps, where a MC step involves a test move of every particle in the system.  

\begin{figure}
\begin{center}
\includegraphics[angle=0,width=8cm]{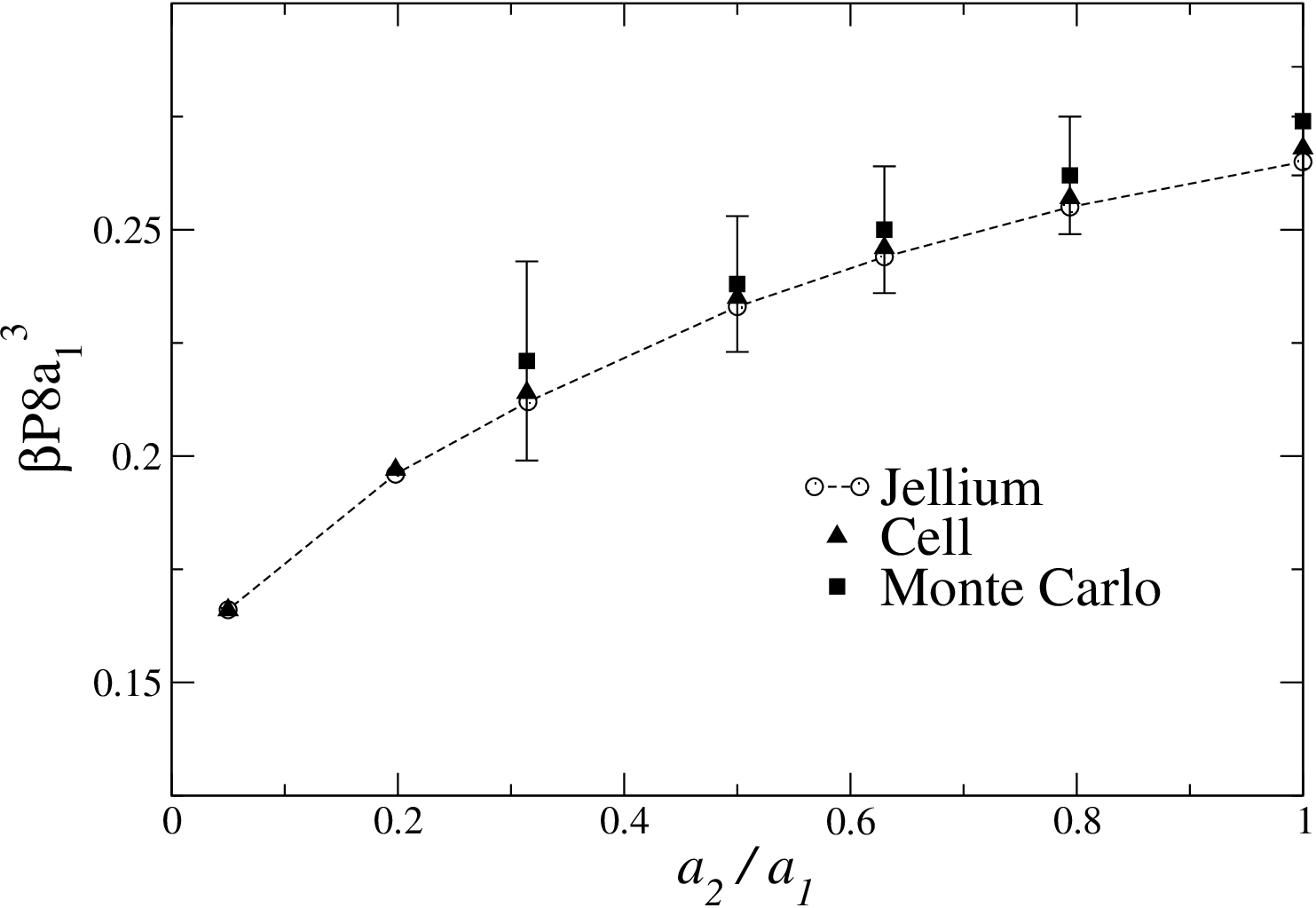}
\end{center}
\caption{Osmotic pressure for a system consisting of two kinds of colloidal particles with the same charge $\wt Z_\bare^i=6.4$, as a function of $a_2/a_1$.
}\label{MC1}
\end{figure}

\begin{figure}
\begin{minipage}{0.48\linewidth}
\begin{center}
\includegraphics[angle=0,width=8cm]{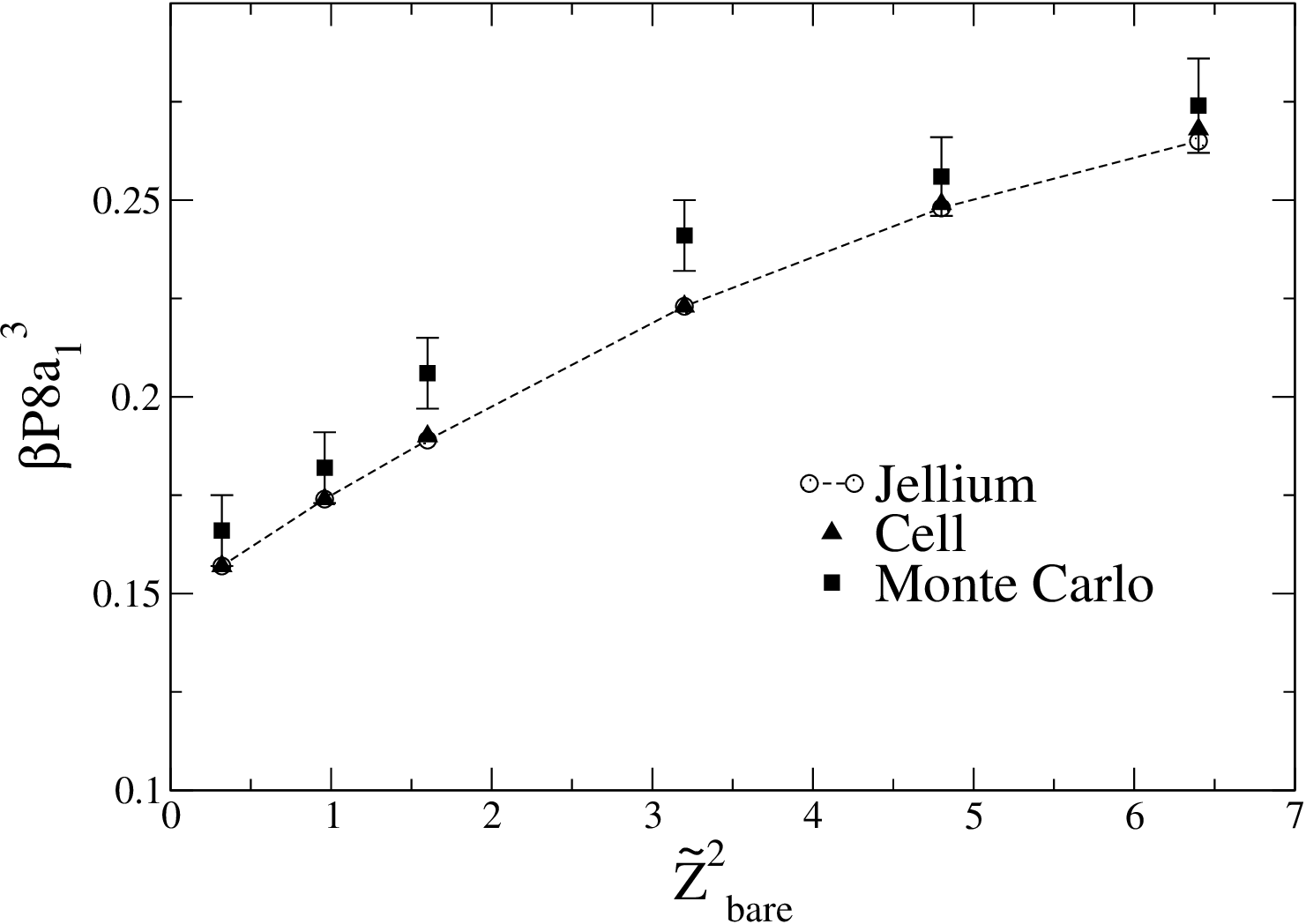}
\end{center}
\end{minipage}
\begin{minipage}{0.48\linewidth}
\begin{center}
\includegraphics[angle=0,width=8cm]{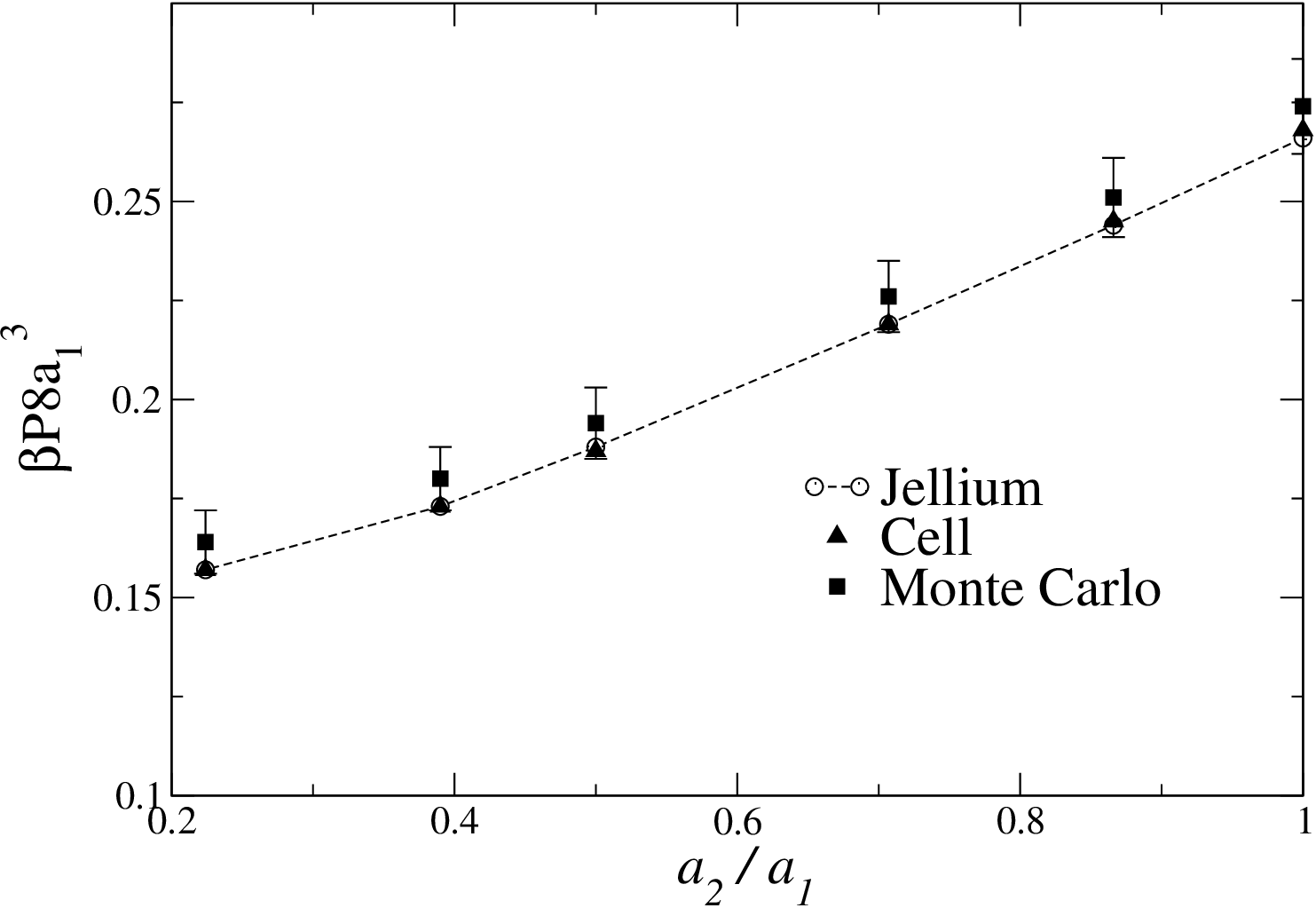}
\end{center}
\end{minipage}
\caption{(Left) Osmotic pressure for a system consisting of two kinds of colloidal particles with the same radius  $a_1=a_2$, as a function of $\wt Z_\bare^2$, with $\wt Z_\bare^1=6.4$. 
(Right) Osmotic pressure, changing the size ratio, keeping a constant surface charge density for both colloids. Here, $a_1$ is fixed, $\wt Z_\bare^i=6.4$, and $a_2$ changes.
}\label{MC2}
\end{figure}

Three sets of simulations were carried out at $\wt \eta=0.01$. In the first set, the two colloidal species have the same bare charge $\wt Z_\bare^1=\wt Z_\bare^2=6.4$ 
(and thus $\Zbare^1=20$), 
while the radius of the second species ($a_2$) is varied. 
We show  in Fig. \ref{MC1}, the simulation results (filled squares) as well as the predictions obtained by the renornalized jellium model (empty circles) 
and the cell model (filled triangles). As can be seen, the agreement between the three sets is very good.
In the second set of simulations, the colloids are all of the same size ($a_1=a_2$), the charge of the first species is kept at $\wt Z_\bare^1=6.4$ and 
the charge of the second species ($\wt Z_\bare^2$) is varied (Fig. \ref{MC2}-left). The results obtained from the jellium and cell models are again nearly 
identical. Although the pressure they predict is in general smaller than that of the MC simulations, the agreement is good. The situation is similar for the 
third set of simulations, (Fig. \ref{MC2}-right) in which  $a_1$ is fixed and $a_2$ varies in such a way as to keep the surface charge density 
($s_i$) constant $s_i=Z_\bare^i/(4\pi a_i^2)$. In all cases, the proximity of cell and jellium 
results is striking, and somewhat surprising given they rely on rather distinct calculations. 

The MC data shown here do not
allow to discriminate one approach against the other. The reason may be that charge renormalization 
effects are not overwhelming with the parameters of the simulations, even if not negligible. 
It would be in this respect 
interesting to increase somewhat the values of the bare charges, to enhance non-linear effects. 
In doing so though, one has to keep in mind that correlation effects will be increased as well,
and when the so called plasma parameter $\Xi_i = 2 \pi \ell_B^2 s_i$ exceeds unity,
the whole Poisson-Boltzmann-like description will start to break down, be it in its
jellium, or in its cell clothing \cite{Levin,SaTr11,rque119}. With the parameters of Fig. \ref{MC2}-right, we have 
$\Xi_1=\Xi_2=1$. On the other hand, with the procedure underlying Fig. \ref{MC1}, we have $\Xi_2 \simeq a_1/a_2$
and therefore, decreasing $a_2/a_1$, Coulombic correlations increase, to reach a value beyond 10 for the
left-most point. In this region, MC simulation are impeded by enhanced equilibration time
(which explains why it is void of MC results).

\section{Conclusions}
\label{sec:concl}

We have proposed a novel procedure for solving jellium-like models,
taking due account of renormalization effects. 
Such approaches had been tested with some success on liposome  and latex dispersions  \cite{hqcseh05,hqcseh06}.
Particular emphasis was put 
on colloidal mixtures, where it was shown that the computationally most
demanding part of the task boils down to a sequence of mono-component 
calculations. The idea was illustrated on binary mixtures, but can be
straightforwardly generalized to arbitrary polydispersities,
including continuous case after suitable discretization.
The method takes advantage of the mean-field nature of the 
theory, where all species considered obey the same Poisson equation,
with different boundary conditions, in a background density
that couples all constituents of the mixture.

In a second step, we have performed Monte Carlo simulations of binary 
mixtures, at primitive model level: the solvent is viewed as a dielectric continuum,
but otherwise, Coulombic interactions are treated exactly. 
This allows to assess the accuracy of mean-field simplifications.
In this respect, we tested the jellium predictions for the osmotic pressure and those of the Poisson-Boltzmann cell,
against Monte Carlo. It was know that in the monocomponent case,
both mean-field approaches yield very close results, that fare very favorably against MC, 
provided of course one remains in the regime
of relatively weak couplings where Poisson-Boltzmann theory may hold.
We have shown here that despite the different nature of the jellium and Poisson-Boltzmann cell approximations, 
both approaches continue to give similar results, close to MC, in the case of binary mixtures of spherical colloids.

Finally, while the method was illustrated on the simplest implementation
of the jellium view, refinements and generalizations along the lines proposed in 
Refs. \cite{C06,Colla1,Colla2} can also be addressed. It is also of interest to
extend our approach to colloidal objects of non spherical shapes.



\begin{thebibliography}{}
\bibitem{rque50}
To this end, the system is enclosed in a spherical large
enough cell of radius $\widetilde R$, such that $\tilde \kappa \widetilde R \gg 1$.
At that point $\tilde r = \widetilde R$, $d\phi/d\tilde r$ is taken to vanish
(electroneutrality),
and the potential $\phi(\widetilde R)$ is varied, in order to 
math a prescribed gradient at $\tilde r=1$, given by the bare charge
(denoted $Y$ here). Under the proviso that the cell is large enough,
the results found do not depend on $\widetilde R$,
which is only used for numerical purposes.
\bibitem{rque10}
Yet, with an asymmetric salt where co-ions have a larger valency than counter-ions, the effective charge is not always smaller than the bare one,
see e.g. Ref. \cite{tt04}.
\bibitem{rque51}
For instance, the parameter $\phi(\widetilde R)$ mentioned in 
\cite{rque50} needs to be varied.
\bibitem{rque60}
The bottom curve with $\kappa_\res a_1=0.5$
exhibits a saturation value below 8 for $\wt Z_\eff^1$, which should also set the upper
possible limit for $\wt Z_\eff^2$, since the two species are of the same size in that 
example. However, the {\it a priori} chosen value was $\wt Z_\eff^2$=9.
\bibitem{rque119}
Yet, it should be noted that $\Xi_i \propto \widetilde Z_i \,\ell_B/a_i$. Thus, with a small enough value of
$\ell_B/a_i$, one can have both a large value of $\widetilde Z_i $ (and thus important non-linear effects),
with a small plasma parameter $\Xi_i$ (validating the mean-field description).
\end{thebibliography}
\end{document}